\documentstyle[12pt,epsf]{article}
\newcommand{\be}{\begin{equation}}
\newcommand{\ee}{\end{equation}}
\newcommand{\bea}{\begin{eqnarray}}
\newcommand{\eea}{\end{eqnarray}}
\newcommand{\bt}{\begin{tabular}}
\newcommand{\et}{\end{tabular}}
\newcommand{\ba}{\begin{array}}
\newcommand{\ea}{\end{array}}
\newcommand{\ov}{\overline}

\begin{document}
\setcounter{page}{0}
\thispagestyle{empty}
\baselineskip=20pt

\hfill{
\begin{tabular}{l}
DSF$-$96/8 \\
INFN$-$NA$-$IV$-$96/8 \\
hep-ph/9607371
\end{tabular}}

\bigskip\bigskip

\begin{center}
\begin{huge}
{\bf On the indistinguishibility of Majorana- from Dirac-neutrino 
propagation in a stellar medium }
\end{huge}
\end{center}

\vspace{2cm}

\begin{center}
{\Large
Salvatore Esposito\\} 
\end{center}

\vspace{0.5truecm}

\normalsize
\begin{center}
{\it
\noindent
Istituto Nazionale di Fisica Nucleare, 
Sezione di Napoli\\
Mostra d'Oltremare Pad. 19-20, I-80125 Napoli Italy\\
e-mail: sesposito@na.infn.it}
\end{center}

\vspace{3truecm}


\begin{abstract}
We study, in the framework of the Standard Model, the propagation 
of (pure) Majorana neutrinos in a typical 
stellar medium and show that Majorana neutrino matter 
oscillations are completely indistinguishable from Dirac ones, 
even if in the case of no family 
mixing Majorana neutrinos can be distinguished from Dirac 
ones in the non-relativistic limit. Moreover, if CP violation is 
present, an effective phase arises in the effective mixing matrix but, 
due to a symmetry of the Majorana fields, this cannot be univocally 
determined
\end{abstract}

\vspace{1truecm}

\newpage

\section{Introduction}

\indent
An open question of the Modern Elementary Particle Physics is 
the problem of neutrino mass. In the Standard Model by Glashow, 
Weinberg and Salam \cite{GWS} neutrinos are massless but there is no 
fundamental reason to believe in that (on the contrary to that happens 
for photons, for which the electromagnetic gauge invariance forces 
them to be massless), and in particular the structure of the Standard 
Model remains unchanged if we introduce a neutrino mass term in the 
electroweak lagrangian. On the other hand, there are some experimental 
indications (even if not conclusive on this argument) \cite{gel} that 
neutrinos could be massive, such as the solar neutrino problem 
\cite{snp}, the atmospheric neutrino anomaly \cite{atmo} and so on. 
Moreover, in Grand Unified Theories it is quite natural for neutrinos 
to acquire a non vanishing mass \cite{GUTs}.

In general, if neutrinos are massive, they could be of the 
``Dirac-type'' or of the ``Majorana-type''. For Dirac neutrinos, in 
strict analogy to that happens for the other fermions present in the 
Standard Model, the particle states are different from the 
anti-particle states, so in addition to $\nu_L$ and $\nu_R^C$, 
revealed by weak interactions, there must be also the states $\nu_R$ 
and $\nu_L^C$, which however are sterile for weak interaction. In a 
lagrangian description, a Dirac neutrino of 4-momentum $k_{\mu}$ 
propagating freely in vacuum is described by
\be
\cal L \; = \; {\ov \nu_L} \not{k} \nu_L \, + \, {\ov \nu_R} \not{k} 
\nu_R \, - \, {\ov \nu_L} M_D \nu_R \, - \, {\ov \nu_R} M_D^{\dagger}
\nu_L
\label{11}
\ee
where $\nu^T \, = \, (\nu_e , \nu_{\mu} , \nu_{\tau})$ are the 
flavour-eigenstate neutrino fields and $M_D$ is a 3x3 Dirac mass 
matrix.

A more economical way for accounting neutrinos mass is that of 
considering neutrinos to be Majorana particles, for which the particle 
states coincide (up to a phase factor) with the anti-particle states, 
so there is no need for introducing sterile $\nu_R$, $\nu_L^C$ states. 
In this case the lagrangian term (\ref{11}) is substituted by
\be
\cal L \; = \; {\ov \nu_L} \not{k} \nu_L \, + \, {\ov \nu_R^C} \not{k} 
\nu_R^C \, - \, {\ov \nu_L} M_M \nu_R^C \, - \, {\ov \nu_R^C} 
M_M^{\dagger} \nu_L
\label{12}
\ee
where now $M_M$ is a 3x3 Majorana mass matrix that, from the fermionic 
anti-commutation rules of the neutrino fields, must be symmetric 
\cite{mp}.

A more general lagrangian is that with both the Dirac mass term in 
(\ref{11}) and the Majorana one in (\ref{12}), whose diagonalization 
leads again to Majorana fields; however, for simplicity, in this paper 
we consider pure Majorana neutrinos with the lagrangian (\ref{12}).

In general, if the mass matrices (Dirac or Majorana) are non diagonal 
(in analogy to that happens in the quark sector) lepton flavour 
violation phenomena can occur, between which neutrino oscillations 
seem to be the most likely to detect experimantally the effects of a 
non vanishing neutrino mass \cite{pontecorvo}.
Anyhow, the properties of Dirac and Majorana neutrinos are completely 
different. For example, in the Majorana case the CPT invariance forces 
neutrinos to have zero diagonal electromagnetic moments, while this 
doesn't happen for Dirac particles. Further, while for Dirac neutrinos 
the global lepton number is a conserved quantity, for Majorana 
neutrinos this is no longer true, and phenomena such as Pontecorvo 
neutrino-antineutrino oscillations \cite{nu-nubar} and neutrinoless 
double beta decay \cite{gel},\cite{mp} can arise. Moreover, if CP invariance 
in the leptonic sector is violated, as in the quark sector, the number 
of physical CP breaking phases in the mixing matrix depends on the Dirac 
or Majorana nature of neutrinos and, in particular, for Majorana 
neutrinos there is one CP breaking phase also in the case of mixing of 
only two families \cite{cp}.

We stress that the knowledge of neutrino properties is of relevant 
importance not only for particle physics, but also for astrophysics 
and cosmology. In fact, for example, the most abundant neutrino 
sources are the active stars, in which neutrinos are produced by 
nuclear processes \cite{astro}; due to their weak interaction, these 
neutrinos play a fundamental role in the mechanism of star cooling. 
However, to this purpose, the study of neutrino interactions with the 
stellar matter (and in particular the study of coherent effects of 
matter on neutrino propagation) is necessary.

In this paper we want to extend the recent analysis \cite{geny} made 
on Dirac neutrino propagation in magnetized matter (such as that 
present in supernov\ae, for example) to the case of Majorana neutrinos 
and to point out the relevant differences of behaviour.

The paper is organized as follows. In section 2 we study the 
propagation of Majorana neutrinos in matter in the case of no family 
mixing for pointing out the main properties. In section 3 the 
coherent effects on neutrino (flavour) oscillations is considered in 
the framework of CP invariance, while in section 4 the insorgence 
of ``effective'' phases in the case of CP violation is studied. 
Finally, in section 5 conclusions are outlined.

\section{Dispersion relation and the free propagation condition}

Let us first consider the propagation of non-mixed Majorana neutrinos 
in matter, so that the mass matrix is diagonal and we can consider one 
neutrino flavour at a time. From the lagrangian in Eq.(\ref{12}) 
follows that in vacuum the equations of motion governing neutrino 
propagation are
\be
\left( \not{k} \, - \, m \right) \, \nu_L \; + \; 
\left( \not{k} \, - \, m \right) \, \nu_R^C \; = \; 0
\label{21}
\ee
When neutrinos propagate in matter, a way for taking into account 
their interactions with the particles in the background is to replace 
the mass $m$ in Eq.({\ref{21}) with the complete self-energy, so the 
equations of motions can be written in the form
\be
\left( \not{k} \, - \, m \, + \, \Sigma_{\nu} \right) \, \nu_L \; + \; 
\left( \not{k} \, - \, m \, + \, \Sigma_{\overline \nu} \right) \, 
\nu_R^C \; = \; 0
\label{22}
\ee
For propagation in a medium with a magnetic field $\vec B$, the 
calculation of $\Sigma_{\nu}$ proceeds with the evaluation of the 
Feynman diagrams in Fig.1. At order $G_F$, the self-energy term 
$\Sigma_{\nu}$ can be cast in the form \cite{geny},\cite{DNPetal}
\be
\Sigma_{\nu} \; = \; b_L \not{u} \; + \; c_L \not{B}
\label{23}
\ee
where $u_{\mu}$ is the medium 4-velocity (for simplicity we consider 
the rest frame of the medium, in which $u_{\mu} \, = \, (1,\vec{0})$ ) 
and $B_{\mu} \, = \, \frac{1}{2} \epsilon_{\mu \nu \alpha \beta} 
u^{\nu} F^{\alpha \beta}$ (in the rest frame $B_{\mu} \, = \, (0, 
\vec{B})$. The coefficients $b_L$,$c_L$ are given by \cite{nr},
\cite{geny}
\be
b_L^{\nu_e} \; \simeq \; - \sqrt{2} G_F ( N_e - \frac{1}{2} N_n )
\label{24}
\ee
\be
b_L^{\nu_{\mu} , \nu_{\tau}} \; \simeq \; \frac{G_F}{\sqrt{2}} N_n
\label{25}
\ee
and
\be
c_L^{\nu_e} \; \simeq \; 0
\label{26}
\ee
\be
c_L^{\nu_{\mu} , \nu_{\tau}} \; \simeq \; - \, \frac{e G_F}{\sqrt{2}} \, 
\frac{(3 \pi^2 N_e)^{\frac{1}
{3}}}{\pi^2}
\label{27}
\ee
for a degenerate Fermi gas, while
\be
c_L^{\nu_e} \; \simeq \; - \, \frac{3e G_F}{4 \sqrt{2}} \, \frac{N_e}{m_e^2}
\label{2.8}
\ee
\be
c_L^{\nu_{\mu} , \nu_{\tau}} \; \simeq \; + \, \frac{3e G_F}{4 \sqrt{2}} \, 
\frac{N_e}{m_e^2}
\label{29}
\ee
for a classical non-relativistic plasma. Here $N_e$,$N_n$ are the 
number density of electrons and neutrons respectively, and we have 
assumed the medium to be electrically neutral.

The Feynman diagrams for $\Sigma_{\overline \nu}$ are showed in Fig.2; 
from \cite{geny},\cite{DNPetal},\cite{nr} it is easy to deduce 
that $\Sigma_{\overline \nu}$ takes the form
\be
\Sigma_{\overline \nu} \; = \;  - \; b_L \, \not{u} \; - \; c_L \, \not{B}
\label{210}
\ee
where $b_L$,$c_L$ are again given by Eqs.(\ref{24})-(\ref{29})
\footnote{$\Sigma_{\overline \nu}$ can be obtained from $\Sigma_{\nu}$
multiplying every neutrinic quantity appearing in that by its phase 
acquired under charge coniugation (in particular $\gamma_{\mu} \, L 
\rightarrow \, - \, \gamma_{\mu} \, R$) and with the substitution $k_{\mu} 
\rightarrow \, - k_{\mu}$. Anyhow, at order $G_F$, the self-energy does 
not depend explicitely on $k_{\mu}$, so that $b_L$,$c_L$ remain 
unchanged.}
Then, the equation of motion can be written as
\bea
\left( \omega \; + \; b_L \; + \vec{\sigma} \cdot \left( \vec{k} \; + 
\; c_L \, \vec{B} \right) \right) \; \nu_L \; & = & \; m \, \nu^C_R  
\label{211} \\
\vspace{1cm}
\left( \omega \; - \; b_L \; - \vec{\sigma} \cdot \left( \vec{k} \; - 
\; c_L \, \vec{B} \right) \right) \; \nu^C_R \; & = & \; m \, \nu_L  
\label{212}
\eea
where $\omega$ and $\vec{k}$ are respectively the neutrino energy and 
momentum, while $\sigma_i$ are the Pauli matrices. Introducing the 
helicity eigenstates $\phi_{\lambda}$ \cite{dolivonieves}
\be
\frac{\vec{\sigma} \cdot  \vec{k}}{| \vec{k} |} \; \phi_{\lambda} \; = 
\; \lambda \, \phi_{\lambda}
\label{213}
\ee
with $\lambda = \pm 1$, because the interaction here considered 
between neutrinos and the particles in the medium preserves neutrino 
helicity \cite{geny}, we can write $\nu_L \, = \, \eta_1 
\,\phi_{\lambda}$ and $\nu_R^C \, = \, \eta_2 \,\phi_{\lambda}$ with 
$\eta_1,\eta_2$ satisfying
\be
\left( \omega \, + \, b_L \, + \lambda k \, + \lambda
c_L \, \frac{\vec{k} \cdot \vec{B}}{k} \, - \, m 
\left( \omega \, - \, b_L \, - \lambda  k \, + \lambda
c_L \, \frac{\vec{k} \cdot \vec{B}}{k} 
\right)^{-1} m \right) \; \eta_1 \; \simeq \; 0 
\label{214}
\ee

\be
\eta_2 \; \simeq \; \left( \omega \; - \; b_L \; - \lambda \, 
\left( k \; - \; c_L \, \frac{\vec{k} \cdot \vec{B}}{k} \right)
\right)^{-1} m \; \eta_1
\label{215}
\ee
From (\ref{214}) we then deduce that for $\lambda \, = \, - 1$ the 
dispersion relation reads
\be
\omega^2 \, - \, k^2 \; = \; m^2 \, - \, 2 k \, b_L 
\, + \, 2 \omega \, c_L  \frac{\vec{k} \cdot \vec{B}}{k}
\label{216}
\ee
while for $\lambda \, = \, + 1$ the signs of $b_L$,$c_L$ are reversed. 
The solution of Eq. (\ref{216}) are the eigen-energies of propagating 
neutrino eigen-modes.

Let us now compare this result with the analogous dispersion
relation found for Dirac neutrinos \cite{geny}:
\be
\omega^2 \, - \, k^2 \; = \; m^2 \, - \, b_L (\omega + k)
\, + \, c_L (\omega + k) \frac{\vec{k} \cdot \vec{B}}{k}
\label{217}
\ee
From this, it is immediately evident that the two dispersion relations 
become equal for ultrarelativistic neutrinos. We stress, however, that 
this conclusion is valid only at order $G_F$. To stop at this order in 
the Fermi coupling constant is sufficient in all physical situations 
in which the background is CP-asymmetric \cite{nr}, such as stellar 
interior. As a consequence, there is no physical difference in the 
behaviour of Dirac and Majorana neutrinos produced in active stars and 
interacting with the matter inside. 
\footnote{In the framework of the Standard Model the intrinsic 
magnetic moment interaction of Dirac and Majorana neutrinos are 
different, but extremely small \cite{mp} compared to the effects here 
considered}
We also observe that for 
non-relativistic neutrinos the obtained results do not apply to 
extremely low energy particles such as relic neutrinos present in our 
epoch. In fact, in this case the neutrino momentum $\vec{k}$ (of the 
order of $10^{-2} \div 10^{-4} \, eV$ for relic neutrinos \cite{kolb})
is no longer conserved in the interaction with the medium, so that the 
same dispersion relation formalism loses of sense.

Let us note that when higher order than $G_F$ become important, as for 
example in the Early Universe, where the particle-antiparticle 
symmetry is believed to be present, the propagation of Dirac neutrinos 
would be quantitatively different from that of Majorana neutrinos.

From (\ref{216}) we also derive the modifications to the free 
propagation condition \cite{geny}
\be
b_L \, k \; = \; c_L \, \omega \, \frac{\vec{k} \cdot \vec{B}}{k}
\label{218}
\ee
If this relation is realized (see \cite{geny} for the conditions under 
which this happens for ultrarelativistic neutrinos) the coherent 
effects dur properly to the magnetic field are opposite to those 
generated in the absence of the field, so that Majorana neutrinos can 
propagate freely in matter as in vacuum.

\section{CP conserving Majorana neutrino matter oscillations}

\indent
Let us now examine the case in which the Majorana mass matrix in (\ref{12}) 
is non-diagonal (but real), so that neutrino flavour oscillations can 
occur; for simplicity, we consider the mixing between only two 
generations (for example $e$,$\mu$).

Now, the eigen-energies of $\lambda = -1$ neutrinos are the solution
of the eigenvalue equation
\be
det \, \left( \omega \; + \; b_L \; - \, k \; - 
\; c_L \, \frac{\vec{k} \cdot \vec{B}}{k} - \; m \,
\left( \omega \; - \; b_L \; + \, k \; - 
\; c_L \, \frac{\vec{k} \cdot \vec{B}}{k} 
\right)^{-1} \, m \right) \; = \; 0
\label{31}
\ee
where $b_L$,$c_L$,$m$ are matrices in the flavour space that, in the 
mass eigenstate basis, have the form \cite{geny}
\be
m \; = \; \left(  \ba{cc}
                               m_1 &  0 \\
                               0  &  m_2
                               \ea  \right)
\label{32}
\ee

\be
b_L \; = \; \left( \ba{cc}
b_L^W \,cos^2 \,\theta \,+ \, b_L^Z & - \,b_L^W \,sin \,\theta \,cos \,\theta\\
- \,b_L^W \,sin \,\theta \,cos \,\theta & b_L^W \,sin^2 \,\theta \, + \,b_L^Z
                 \ea  \right) 
\label{33}
\ee

\be
c_L \; = \; \left( \ba{cc}
c_L^W \,cos^2 \,\theta \,+ \, c_L^Z & - \,c_L^W \,sin \,\theta \,cos \,\theta\\
- \,c_L^W \,sin \,\theta \,cos \,\theta & c_L^W \,sin^2 \,\theta \, + \,c_L^Z
                 \ea  \right) 
\label{34}
\ee
Here, $\theta$ is the vacuum mixing angle while $b_L^W$,$c_L^W$ and
$b_L^Z$,$c_L^Z$ are respectively the charged-current and 
neutral-current contribution to the coefficients in the self-energy. 
After simple algebra, in the approximation of non-exceptionally dense 
medium, we arrive at the following eigenvalue equation
\footnote{Neglecting damping effects, the energy eigenvalues 
completely determine neutrino flavor 
oscillations. In fact, all the physically interesting quantities 
(survival probability, resonance and no-resonance conditions and so 
on) can be expressed in terms of these quantities, as can be 
seen, for example, in \cite{geny}}
\bea
\left( \omega \, - \, \ov{k_1} \, - \, \frac{m_1^2}{\omega \, + \, k_1} 
\right) \, 
\left( \omega \, - \, \ov{k_2} \, - \, \frac{m_2^2}{\omega \, + \, k_2} 
\right) \; +  
\nonumber
\eea
\bea
+ \; (\omega \, - \, \ov{k_1}) (\omega \, - \, \ov{k_2}) \,
\left( \left( 1 \, - \, \frac{S_{12}^2}{
(\omega \, + \, k_1) (\omega \, + \, k_2)} \right) \, 
\left( 1 \, - \, \frac{D_{12}^2}{
(\omega \, - \, \ov{k_1}) (\omega \, - \, \ov{k_2})} \right) \; 
- \; 1 \right) \; + 
\nonumber
\eea
\be
+ \; \frac{2 \, m_1 \, m_2 \, S_{12} \, D_{12}}{
(\omega \, + \, k_1) (\omega \, + \, k_2)} \;  = \; 0 
\label{35}
\ee
where the matrices $S,D$ are given respectively by
$b_L \, + \, c_L \, \frac{\vec{k} \cdot \vec{B}}{k}$, 
$b_L \, - \, c_L \, \frac{\vec{k} \cdot \vec{B}}{k}$ and 
$k_1 \, = \, k - S_{11}$, $k_2 \, = \, k - S_{22}$, 
$\ov{k_1} \, = \, k - D_{11}$, $\ov{k_2} \, = \, k - D_{22}$. Compare 
this result with that analogous for Dirac neutrinos \cite{geny}
\be
\left( \omega \, - \, \ov{k_1} \, - \, \frac{m_1^2}{\omega \, + \, k} 
\right) \, 
\left( \omega \, - \, \ov{k_2} \, - \, \frac{m_2^2}{\omega \, + \, k} 
\right) \; - \; D_{12}^2 \; = \; 0
\label{36}
\ee
Let us observe that in astrophysical environments the terms
\be
\frac{b_L^W \, - \, c_L^W \, \frac{\vec{k} \cdot \vec{B}}{k}}{\omega 
\, + k} \;\;\;\;\;\;\; , \;\;\;\;\;\;\;
\frac{b_L^Z \, - \, c_L^Z \, \frac{\vec{k} \cdot \vec{B}}{k}}{\omega 
\, + k}
\label{37}
\ee
are completely negligible, so that we arrive at the conclusion that 
Majorana neutrino matter oscillations are indistinguishable from Dirac 
ones, irrispective of the relativistic properties of the neutrinos 
themselves. Obviously, as in section 2, this conclusion is not 
applyable to oscillations in the Early Universe.

\section{Effective CP violating phases}

\indent
Finally, let us consider the general case of a complex non-diagonal 
Majorana mass matrix arising when CP invariance in the leptonic 
sector is violated. Now, the mixing matrix $V$ is unitary and, in 
general, for two generations three phases are present, two of which 
physical meaningless because they can be absorbed \cite{mp} by the 
charged lepton fields in the weak charged interaction lagrangian term. 
Then the mixing matrix $V$ can be cast in the form \cite{cp}
\be
V \; = \; \left( \ba{cc}
                 cos \, \theta & - \, sin \, \theta \, e^{i \delta} \\
                 sin \, \theta \, e^{-i \delta} & cos \, \theta
                 \ea  \right)
\label{41}
\ee
Considering matter oscillations, the eigenvalue equation is again of 
the form (\ref{31}), but now the matrix $b_L$, in the mass eigenstate 
basis, is given by
\be
b_L \; = \; \left( \ba{cc}
b_L^W \,cos^2 \,\theta \,+ \, b_L^Z & - \,b_L^W \,sin \,\theta 
\,cos \,\theta \, e^{i \delta}\\
- \,b_L^W \,sin \,\theta \,cos \,\theta \, e^{-i \delta} & 
b_L^W \,sin^2 \,\theta \, + \,b_L^Z
                 \ea  \right) 
\label{42}
\ee
and analogously for $c_L$. For definiteness, let us consider only 
ultrarelativistic neutrinos. The eigenvalue problem (\ref{31}) is 
equivalent to the diagonalization of the hamiltonian 
\cite{dolivonieves}
\be
H \; \simeq \; k \, + \, \frac{m^2}{2k} \, - \, b_L \, + \, c_L 
\, \frac{\vec{k} \cdot \vec{B}}{k}
\label{43}
\ee
that, disregarding terms proportional to the identity matrix which 
play no role in neutrino oscillations, in the flavour 
eigenstate basis takes the form
\be
H_{flav} \; = \; \frac{\Delta m^2}{2k} \;
\left( \ba{cc}
sin^2 \,\theta  & - sin \,\theta \,cos \,\theta \, e^{i \delta}\\
- \,sin \,\theta \,cos \,\theta \, e^{-i \delta} & cos^2 \,\theta 
       \ea  \right) \; - \; 
\left( b_L^W \, - \, c_L^W \, \frac{\vec{k} \cdot \vec{B}}{k} \right)
\left( \ba{cc}
1 & 0 \\
0 & 0
\ea \right)
\label{44}
\ee
This is an hermitian matrix that can be diagonalized by a unitary 
effective mixing matrix $\tilde{V_m}$ ( $\tilde{V_m^{\dagger}} 
\, H_{flav} \, \tilde{V_m} \, = \, H_m$ with $H_m$ diagonal and 
real) whose general form is \cite{cp}
\be
\tilde{V_m} \; = \; P \, V_m \; = \; 
\left( \ba{cc}
              e^{i \theta_1} & 0\\
              0 & e^{i \theta_2}
       \ea \right) \, 
\left( \ba{cc}
        cos \, \theta_m & - \, sin \, \theta_m \, e^{i \alpha} \\
        sin \, \theta_m \, e^{-i \alpha} & cos \, \theta_m
                 \ea  \right)
\label{45}
\ee
Obviously, the phases $\theta_1$,$\theta_2$ don't produce separately 
observable effects, but only their difference $\beta \, = \, \theta_2 
\, - \, \theta_1$ could. In fact, $P^{\dagger} \, H_{flav} \, P$ has 
the same form as $H_{flav}$ but with $\delta$ replaced by $\delta \, + 
\, \beta$. Furthermore, regarding neutrino oscillations, only one 
effective phase $\delta_m$ plays a physical role. In fact, the 
diagonalization equation which determine the effective mixing angle 
$\theta_m$ and the effective phase $\delta_m$ are
\bea
\left( \frac{\Delta m^2}{2k} \, cos \, 2 \theta + \; 
b_L^W \, - \, c_L^W \, \frac{\vec{k} \cdot \vec{B}}{k} \right) \,
sin \, 2 \theta_m \; = 
\nonumber
\eea
\be
= \; \frac{\Delta m^2}{2k} \, sin \, 2 \theta \,
\left( cos^2 \, \theta_m \, e^{i (\delta \, + \, \delta_m)} \, - \,
sin^2 \, \theta_m \, e^{-i (\delta \, + \, \delta_m)} \right)
\label{46}
\ee

\be
cos^2 \, \theta_m \, e^{i (\delta \, + \, \delta_m)} \, - \,
sin^2 \, \theta_m \, e^{-i (\delta \, + \, \delta_m)} \; = \;
cos^2 \, \theta_m \, e^{-i (\delta \, + \, \delta_m)} \, - \,
sin^2 \, \theta_m \, e^{i (\delta \, + \, \delta_m)}
\label{47}
\ee
with $\delta_m \, = \, \beta \, - \, \alpha$. From these we get that 
the effective mixing angle $\theta_m$ is unaffected respect to the 
case of CP conservation, and this follows from the fact that the 
effective phase is exactly opposite to the intrinsic (vacuum) one:
\be
tg \, 2 \theta_m \; = \; \frac{ \frac{\Delta m^2}{2k} \, sin \, 2 
\theta}{\frac{\Delta m^2}{2k} \, cos \, 2 
\theta \; + \; b_L^W \, - \, c_L^W \frac{\vec{k} \cdot \vec{B}}{k} }
\label{48}
\ee

\be
\delta_m \; = \; - \, \delta
\label{49}
\ee
We now give a physical interpretation of the obtained results.

In vacuum, there is no way to observe CP violating phases in flavour 
oscillations, because a CP violation means that there is  a 
difference in the physical behaviour of neutrinos and antineutrinos, 
so that if we observe only transistions between neutrinos (even if of 
different flavour) this difference can not be detectable. The same is 
true for flavour matter oscillations (even if neutrinos and 
antineutrinos have different refractive index), so the effective
mixing angle $\theta_m$ must have no changes and this explains (\ref{48}).

Note, however, that (\ref{49}) is a bit more general condition, and it is 
surprising the fact that the effective CP violating phase $\delta_m$ 
determined by (\ref{49}) does not determine univocally the effective phase 
$\alpha$ present in the mixing matrix (\ref{45}). This result is a 
consequence of the fact that the physical meaningful effective mixing 
matrix has the same form as the vacuum mixing matrix, i.e. it is 
$V_m$, not $\tilde{V_m}$, or, in other words, that the matrix $P$ 
has no physical effect. In fact, as already noted, it has no effect on 
the hamiltonian ($\delta \rightarrow \delta \, + \, \beta$), but this 
is not sufficient because it must have no effect also on the Majorana 
matter mass eigenstate neutrino fields. That happens because these 
fields can be written in the form
\be
\left( \ba{c}
         \nu_{1m}\\
         \nu_{2m}
       \ea \right) \; = \;  V_m^{\dagger} \, P^{\dagger} \, 
\left( \ba{c}
         \nu_{eL}\\
         \nu_{\mu L}
       \ea \right) \; + \;  V_m^T \, P \,
\left( \ba{c}
         \nu_{eR}^C\\
         \nu_{\mu R}^C
       \ea \right) 
\label{410}
\ee
and the phases present in $P$ can be absorbed $simultaneously$ by 
$\nu_L$ and $\nu^C_R$, redefining the physical states as
\be
\left( \ba{c}
         \nu_{eL}\\
         \nu_{\mu L}
       \ea \right) \; \rightarrow \; P^{\dagger} \, 
\left( \ba{c}
         \nu_{eL}\\
         \nu_{\mu L}
       \ea \right) \; = \; 
\left( \ba{c}
         e^{-i \theta_1} \, \nu_{eL}\\
         e^{-i \theta_2} \, \nu_{\mu L}
       \ea \right) \;
\label{411}
\ee
\be
\left( \ba{c}
         \nu_{eR}^C\\
         \nu_{\mu R}^C
       \ea \right) \; \rightarrow \; P \, 
\left( \ba{c}
         \nu_{eR}^C\\
         \nu_{\mu R}^C
       \ea \right) \; = \; 
\left( \ba{c}
         e^{i \theta_1} \, \nu_{eR}^C\\
         e^{i \theta_2} \, \nu_{\mu R}^C
       \ea \right) \;
\label{412} .
\ee

\section{Conclusions}

\indent
In this paper we have analyzed the behaviour of pure Majorana neutrinos
when they pass through a dense medium (typically a stellar medium) and 
outlined the main differences with the behaviour of Dirac neutrinos. 

The self-energy in matter of neutrino and antineutrino states are 
different (compare (\ref{23}) with (\ref{210}) ). However, this brings to 
observable differences in the propagation of Majorana neutrinos in 
stellar matter with respect to the Dirac case only for 
non-relativistic neutrinos (at first order in the Fermi coupling 
constant). It is anyhow remarkable the fact that Majorana neutrino 
matter oscillations are completely indistinguishable from Dirac ones 
in astrophysical environments (both for ultrarelativistic anf for 
non-relativistic neutrinos), due to the smallness of the effective 
potential (\ref{37}).

Finally, we have considered the case of a possible CP (intrinsic) 
violation that can be present, for Majorana neutrinos, also in the 
mixing of only two generations. Obviously this has no effect on 
neutrino flavour conversion (both in vacuum and in matter) but should 
have on neutrino-antineutrino oscillations. We have then analyzed the 
properties of the effective unitary mixing matrix and have shown that 
the effective phase $\alpha$ present in it, due to the invariance for 
phase change of the Majorana matter eigenstate neutrino fields
(\ref{410}), is not univocally determined.

\vspace{1truecm}

\noindent
{\Large \bf Acknowledgements}\\
\noindent
The author is greatly indebted with Prof. F.Buccella for very 
stimulating discussions and valuable hints.

\end{document}